# Community-scale Big Data Reveals Disparate Impacts of the Texas Winter Storm of 2021 and its Managed Power Outage

Cheng-Chun (Barry) Lee[a]*, Mikel Maron[b], and Ali Mostafavi[c]

[a] Postdoctoral Research Associate, Zachry Department of Civil and Environmental Engineering, Texas A&M University, 199 Spence St., College Station, TX 77840; e-mail: ccbarrylee@tamu.edu

[b] Community Team, Mapbox, Washington, DC; e-mail: mikel@mapbox.com

[c] Associate Professor, Zachry Department of Civil and Environmental Engineering, Texas A&M University, 199 Spence St., College Station, TX 77840; e-mail: amostafavi@civil.tamu.edu

## Abstract

Aggregated community-scale data could be harnessed to allow insights into the disparate impacts of managed power outages, burst pipes, and food inaccessibility during extreme weather events. During Winter Storm Uri in February 2021, Texas power-generating plant operators resorted to rolling blackouts to prevent wider collapse of the whole power grid when power demand overwhelmed supply. In this study, we collected community-scale big data (e.g., digital trace and crowdsourced data) in the context of the winter storm, which brought historically low temperatures, snow, and ice to the entire state of Texas. By analyzing this data, we can infer the extent of disparities in storm-related impacts on vulnerable populations. In addition to assessing disparate impacts of the managed power outage, this study also examined the extent of burst pipes and disrupted access to food. Using statistical and trend classification analyses, the results highlight the spatial and temporal patterns of impacts for subpopulations in Harris County, Texas, within which the city of Houston is located. In particular, findings show significant disparity in the extent and duration of power outages experienced by low-income and minority groups. This finding suggests that the existence of inequality in the management and implementation of the power outage. Also, the extent of burst pipes and disrupted food access were more severe for low-income and minority groups. These findings suggest implications for understanding the disparate impacts of extreme weather events on vulnerable populations; in particular, the findings could provide insights allowing infrastructure operators to enhance social equality considerations during managed service disruptions in such events. The results and findings demonstrate the value of community-scale big data sources for rapid impact assessment in the aftermath of extreme weather events.

# Introduction

Extreme weather disasters, such as floods, hurricanes, earthquakes, and winter storms are likely to become more frequent and more intense, leading to more life-threatening and economy-damaging events. (*1–3*) Such events could cause severe damage of critical infrastructure systems (e.g., power, water, and road networks), which will inflict severe impacts on human systems and the environment. (*4–7*) Compounding infrastructure failures, socioeconomic and race inequality can cause different levels of vulnerability to disasters. (*8–10*) Understanding the disparate impact of the past events on minorities becomes critical when developing mitigation plans to reduce the influence of extreme weather events on society and to prepare response strategies to adapt to other disasters. (*2, 11*)

Communities are not monolithic entities, meaning that people's awareness, needs, responses, and tolerance to a disaster can vary greatly. It has been known and shown in the literature that inequalities in race, ethnicity, income, gender, and age exist during and in the aftermath of disasters (12–15). In particular, extreme weather disasters can result in service disruptions in critical infrastructures, such as transportation, electricity, water, and communication systems (16). Elliott and Pais (13) examined the impacts of Hurricane Katrina by analyzing survey data collected from more than 1,200 survivors and found strong differences in race and class. Peacock et al. (15) used tax appraisal data to assess long-term trends in housing recovery for Hurricanes Andrew and Ike and showed that low-income areas tend to suffer more damage and recover more slowly. Coleman et al. (17) investigated empirical data after Hurricane Harvey to assess the impacts of service disruptions on critical infrastructures. The results revealed that less advantaged socioeconomic households, racial minorities, and households with younger residents reported less ability to withstand disruptions. In sum, the literature has revealed existing inequalities in the context of disaster management. Studies on extreme cold weather situations, however, are scarce. The Texas winter storm, Winter Storm Uri, is a unique example affording the opportunity to examine to what extent the impacts of power outages, burst pipes, and food inaccessibility were disproportional across different races, ethnicities, and incomes.

Due to climate change, the frequency and intensity of extreme weather events have been increasing. Such extreme weather may necessitate that power infrastructure owners and operators implement managed outages to cope with the surge in demand in the case of both winter storms and extreme heat or to prevent wildfires to start. Despite the growing practice of managed power outages during extreme weather events, little evidence exists regarding the extent to which these managed outages are implemented in an equitable manner. This limitation is partly due to the inability of researchers to access fine resolution data related to the extent and duration of outages for different subpopulations. In this study, we address this limitation by harnessing and analyzing digital trace data to obtain proxy measures to analyze the extent of power outages experienced by subpopulations. In addition to the use of digital trace data for examining the extent of power outages, we also utilized other community-scale data sources for rapid impacts assessment related to burst pipes and disrupted access to food.

Several methods have been proposed to evaluate the community impact of crisis events using new technology that provides large-scale digital trace data. (*10, 18*) Detailed spatiotemporal data offer unique insights into the interdependence of disaster problems, human activity, and urban mobility. The new insights can help us assess the impact of disaster more accurately (*19*), respond to disaster promptly (*20*), and enhance preparedness for future events (*21*). Researchers have relied mainly upon social media platforms to evaluate disaster impact in recent years. For example, social media has been implemented to sense community disruptions impact (*19*), assess disaster footprints and damages (*22*), categorize disasters for responses (*23*), and map disaster locations (*24*). Social media data, however, can be affected by factors such as income, population size, population composition, or minority ratio. (*25*) The evaluation results from social media may be potentially biased because of the unbalanced user population among socioeconomic groups and affected areas. In addition, not every piece of social media information is geo-coded; for example, only 1% to 2% of tweets have geospatial information (*26, 27*), which may lead to biased results and stymies tracking population activity and behavior difficult.

In this study, we utilized digital trace and crowdsourced datasets to evaluate the impact of the 2021 winter storm in Texas. Population activity can be derived from cell phone signal densities. Higher cell phone signal densities may indicate more population activities in a location; thus, cell phone signal densities can be aggregated into an activity index as a proxy for population activity. A few studies have incorporated activity index into disaster-related research; for example, Yuan et al. identified human activity features for flood impacts (*30*), and Gao et al. discovered early indicators for COVID-19 (*31*). The literature demonstrates that the analysis of digital trace data enables assessment of the impact of disaster events. In particular, digital trace data could provide a reliable proxy for the extent and duration of power outages. The level of cellphone signal activity would fluctuate if there is power outage. (People may evacuate if they don't have power, or they might not have power to charge their cellphone devices.) This study obtained and integrated digital trace data from several sources to examine the extent and duration of the power outages. In addition to digital trace data based on cellphone activities, the time series of population visits to points of interest (POIs) is a commonly used indicator of population activities. Analysis of visits to POIs, such as restaurants, gas stations, and grocery stores, reveals insights into how people move around a city and where people visit during a disaster event. Several studies have been conducted to assess the impact or determine the signal of disasters, such as the COVID-19 pandemic (*20, 28*), Hurricane Harvey (*18*), and Hurricane Irma (*29*), by implementing the visit data of POIs. The fluctuations in POI visits could provide reliable insights regarding disrupted access to critical facilities, such as grocery stores and restaurants.

By employing community-scale big data from various sources, the objectives of the paper are to (1) categorize the responses to infrastructure failure (e.g., managed power outages) and to assess the impacts (e.g., burst pipes and food inaccessibility) of an unusual extreme weather event, and (2) reveal potential disparate impacts on income, race, and ethnicity in responding to an unusual event. The study used the Texas winter storm, Winter Storm Uri, in February 2021, as a case study to fulfill the objectives of the paper.

## Materials and methods

This study gathered and aggregated data from several data sources related to population activity, point-of-interest visits, 311 service helpline, and demographic information. As shown in Figure 1, the data were collected and used to assess community impacts related to the winter storm-induced power outages, burst pipes, and food inaccessibility. All data were aggregated to the census tract level to be comparable. The study area, period, and details of each data source are discussed in the following sections.

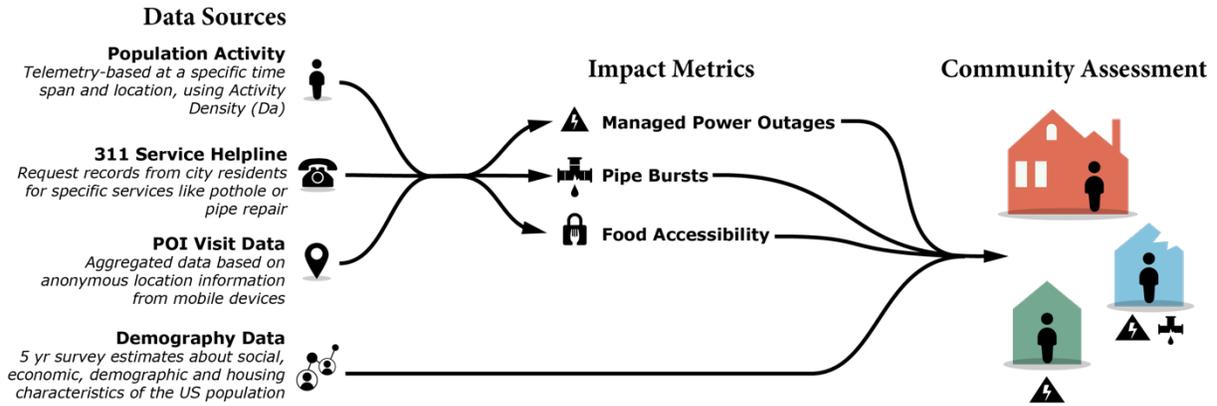

Figure 1 Schematic of the study to assess the impact of Winter Storm Uri.

*Study area and period*

The study collected and analyzed data from various, sources, Harris County, Texas, which includes Houston metropolitan area. From February 13 through 17, 2021, a severe and icy winter storm, Winter Storm Uri, affected most of North America. The Houston area was one of the most adversely affected areas in the state. The winter storm reached Harris County late on the evening of February 14, 2021, bringing snow and below-freezing temperatures. Residents of Harris County were warned of the impending snow and low temperatures starting February 10, 2021. On the afternoon of February 12, 2021, disaster declarations were in effect in all 254 Texas counties. (The National Weather Service issued winter weather alerts to more than 154 million people in the United States on February 15, 2021. (*32*)) The impact of the storm was especially significant on Texas residents who were not experienced with cold weather, snow, or icy conditions. Historic records, both temperature and duration, were broken, and at least 111 people passed away in Texas due to hypothermia. (*33–35*) This winter storm caused severe damage in Texas. Traffic was greatly affected due to icy and slick road conditions. A 133-car pileup on Interstate Highway 35W in Fort Worth caused six fatalities. (*36*) The state's electricity infrastructure was overwhelmed due to heating demand, necessitating rolling blackouts to avoid collapse of the entire electrical grid. Nearly 4.5 million Texas homes and businesses lost power during the peak of this crisis. (*34*) The estimated cost of the blackouts was at least $195 billion USD. (*37*) In the study area of Harris County, more than 90% of residents lost power, and about 65% of the residents experienced water outages at some point during the winter storm. (*38*) For this study, we obtained data from January and February 2021 to conduct the analyses. The January 2021 data were averaged based on the day of the week and served as the baseline for the assessments. The study then used the data in February 2021 to compare with the baseline data to understand the impact of disaster preparedness, responses, and recovery. Data from different sources were aggregated at the census tract level and integrated to be comparable.

*Data sources*

*Population activity*

Due to a lack of publicly accessible power outage records of sufficiently fine granularity, this study relied upon data provided by Mapbox to obtain telemetry-based population activity at a specific time period and location to assess the impact of power outages. Issues during blackout periods would cause a decrease in telemetry-based population activity. For example, telecommunication base stations could not transmit signals, cable modems and Wi-Fi routers in houses could be nonfunctional, cell phones' batteries could be without a means of recharging. Attempts by the authors to obtain fine-resolution power outage data from the electricity owner and operator company were unsuccessful. Therefore, this study assessed the impact of power outages during the winter storm by analyzing telemetry-based population activity. Mapbox aggregated raw cell phone information by time span and geographic unit and normalized them to a baseline period, January 11 to 17, 2021, in this study. After Mapbox aggregated and normalized the raw data, activity index (*A*), a scaling factor, was calculated by dividing the cell phone device counts of geographic units over the 99.9th percentile across all geographic units in the baseline period. A geographic unit is a point in a spatial-resolution grid representing certain areas in the study areas. The 99.9th percentile of cell phone device counts across all geographic units represents the highest population activity of the study areas excluding some potential outliers. In other words, a geographic unit received 0.5 at a specific time span indicating that the cell phone device counts of the geographic unit are 50% of the highest cell phone device counts of all geographic units at its baseline period. The more users located in a geographic unit at a time span, the higher the population activity. Mapbox applied two anonymization processes to protect users' privacy. Firstly, if geographic areas have only a few counts below minimum requirements, the counts would be dropped. Secondly, slight random noise was applied to the existing counts. While the data was derived from cell phone information, data may not exist in every geographic unit at all times. For example, a National Park has many tourists during the day but closes at night, resulting in less population activity and thus insufficient data to be aggregated during nighttime. Some places may lack data if the visit counts are too few at a particular time. The spatial resolution of the aggregated data is about 100 by 100 meters, and the temporal resolution is 4 hours.

To make the data comparable, the activity index was aggregated and used to calculate the activity density (*Da*) of each census tract (*ct*) at each time period (*t*). The calculation is:

$$Da(ct, t) = \sqrt{\frac{1}{N}\sum_{u=1}^{N} A_{u,t}^2}$$

where, *Da(ct, t)* is the activity density at time *t* in census tract *ct*, $A_{u,t}$ is the activity index at time *t* of geographic unit *u*, and *N* is the number of geographic units within census tract *ct*. By calculating activity density for every census tract in Harris County and every time period between February 1 through 28, 2021, the spatiotemporal pattern of population activity can be identified.

*Point-of-interest data*

POI data used in this study was provided by SafeGraph, who partners with several mobile applications to obtain anonymous location data from mobile devices. The aggregated POI data includes basic information of places, such as location name, latitude, and longitude, address, and business category denoted by standard North American Industry Classification System (NAICS) code, which identifies specific business categories. In addition, the POI data provides the visit counts to a POI every day over the analysis period.

To evaluate the impact of the winter storm, this study identifies two types of POIs for analyses: grocery stores (NAICS code: 445110) and restaurants (NAICS code: 72251). The data for grocery stores can represent the changes in terms of daily needs of food accessibility. The fluctuation of restaurant visits can be extended to understand the food accessibility of communities. Visit data was aggregated at the census tract level to isolate the visit changes during the winter storm; however, some residents living close to the boundary of census tracts may tend to go to the restaurants or grocery stores in neighboring census tracts. To account for this consideration, a one-mile buffer was applied to all census tracts when aggregating visit data to the census tract level. In other words, the aggregated visit data of a census tract would include the POIs within a one-mile distance from the boundary of the census tract.

*311 Service Helpline*

The 311 Service Helpline in the Houston area is a consolidated call center at which city residents can report non-emergency concerns, such as traffic fines and sewer concerns to pothole problems and neighborhood complaints and can request information on city services. The details of each 311 record include start and finish time, latitude and longitude, and the type of service request. In this study, the 311 data is used to assess the impact of burst pipes. Thus, water-related service requests were filtered to reveal the 311 records related to burst pipes. The data were further aggregated to the census tract level based on the latitude and longitude of each request. In this manner, the count of cases of burst pipes in each census tract on each day in February 2021 can be calculated and used for analysis.

*Demography*

All demographic and household socioeconomic data used in this study were retrieved from the American Community Survey database administrated by US Census Bureau and aggregated by census tract level. The year and the version of the data are the 2019 5-year estimates data, representing the estimates over the five-year period from 2015 to 2019. The census tracts within the top or bottom 25% of data were classified as the high- and low-income and the racial/ethnic minority and nonminority groups. (Figure 2) The low-income group represents the census tracts having relatively low median household income. The racial minority group represents the census tracts having a relatively large percentage of Black or African American population. The ethnic minority group represents the census tracts having a relatively high ratio of persons of Hispanic origin. The demographic data thus obtained can be compared with the assessment of the impact of the winter storm and used to understand the disparity in experience between various income, racial, and ethnical groups during the disaster.

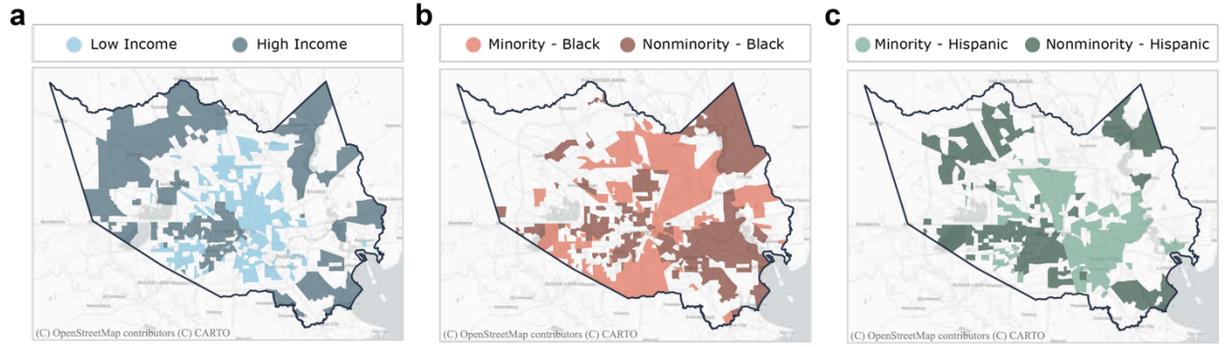

Figure 2 Locations of the high and low-income groups and racial/ethnic minority and nonminority groups. High/low income (a) and minority/nonminority groups for Black (b) and Hispanic (c) populations were identified in terms of the top 25% and bottom 25% levels, respectively, in Harris County.

*Trend classification*

Besides the statistical analysis and comparisons of spatial patterns, an agglomerative hierarchical clustering algorithm (*39*) was implemented to classify neighborhood visit trends and also to understand the patterns response by residents of a census tract to the impact of the winter storm. The classification of POI visitation trends enables us to evaluate the variation of impacts across subpopulations. Agglomerative clustering is a commonly used hierarchical clustering method to group data points into clusters based on their similarity or distance. The algorithm starts by treating each data point as a single cluster. At each iteration, the distances between every pair of clusters in the dataset are calculated. Also, similar clusters are merged with neighboring clusters based on the proximity until all clusters are combined into one big cluster containing all data points. The study implements Euclidean distance to determine similarity of different clusters and uses Ward's linkage criterion, which minimizes variance within clusters, to determine the clusters to be merged at each iteration. In particular, Ward's linkage criterion is described as the following expressions:

$$Error\ Sum\ of\ Squares\ (ESS) = \sum_{i=1}^{n} \left| x_i - \frac{1}{n}\sum_{j=1}^{n} x_j \right|^2$$

$$D(X,Y) = ESS(XY) - [ESS(X) + ESS(Y)]$$

where *X* and *Y* are two clusters, *XY* is the combined cluster of *X* and *Y*, and *n* is the number of data points in *x* cluster.

## Results

After data processing and aggregation of all data at the census tract level, we assessed community impacts due to the Texas winter storm in three aspects: power outages, burst pipes, and food inaccessibility. The impact of power outages caused by the winter storm was assessed based on population activity data in the absence of granular power outage data. During the extreme cold weather, the electricity demand increased, while the supply decreased significantly due to the vulnerable power-generating facilities in Texas that were not winterized nor designed to operate in unseasonable cold. (*40*) The Electric Reliability Council of Texas (ERCOT) had no choice but to shut down the partial power grid to avoid complete failure. It is important to understand, however, whether these outages were experienced in an equitable manner by different subpopulations in terms of race, ethnicity, and income. In addition, this study assessed the impact of pipe bursts according to 311 calls in Harris County related to water damage. Due to the cold weather and limited preparedness beforehand, reports of frozen and burst water pipes were widespread in Harris County. Food accessibility was also limited and critical during the winter storm because of the traffic conditions. Visits to restaurants and grocery stores during the winter storm were used to assess the impact of the winter storm on food accessibility. Through this analysis, results based on community-scale big data from various resources were used to address two research objectives: (1) to quantify and to evaluate the spatial patterns of the impacts and responses of the winter storm on the community; (2) to evaluate the extent of disparities in impacts experienced by low-income and racial/ethnic minority subpopulations. The following sections will present and discuss the results in more detail.

*Impact of Power Outages*

Without electricity, the functioning of most appliances, as well as communications, water, and transportation may be disrupted. Thus, assessing the impact of power outages and understanding which populations and areas are affected is critical for orderly disaster recovery; however, despite efforts by the authors to gather fine-grained and high-resolution power outage data to enable impact and disparity assessment, publicly accessible power outage records are not granular enough to be used for this purpose. Thus, telemetry-based population activity data served as a reliable proxy for examining the extent of power outages. A consequence of a power outage may be loss of internet connectivity or persons traveling from their houses to find a power source in an unaffected area. Activity index, a telemetry-based index provided by Mapbox, was aggregated and used to calculate activity density ($D_a$) at the census tract level. More than half of the census tracts in Harris County, however, did not register enough activity to calculate activity density during the night. Only the data between 8:00 a.m. and 8:00 p.m. were included for analysis. Figure 3 demonstrates the activity density changes of a census tract in Harris County compared to its baseline period. The census tracts experienced an almost 50% decrease in population activity on February 15, 2021 then recovered to the normal state with a few fluctuations about February 22, 2021. According to the experience of a resident living in this census tract during the winter storm, the power outage began on February 15, 2021, which coincided with the greatest dip shown in Figure 3. The power started to recover intermittently from February 17 to February 19, 2021, which matched the fluctuations. On and after February 20, 2021, the power was fully recovered and the activity level began to return to normal, evident in the significant jump in activity density between February 19 and 20, 2021.

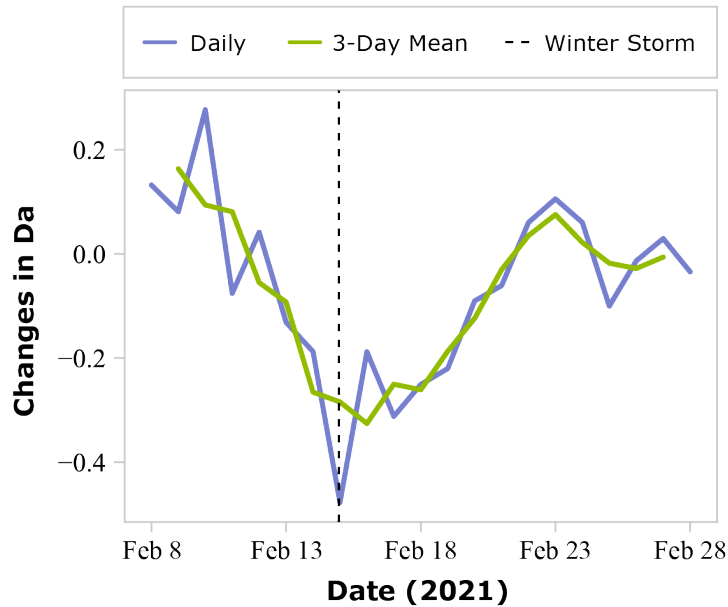

Figure 3 Changes in activity density of a census tract in Harris County, which experienced an almost 50% decrease in population activity on February 15, 2021, when the winter storm reached, and recovered to the normal states with a few fluctuations about February 22, 2021.

In this study, two features of activity density, the greatest negative changes and recovery duration, were examined to assess the impact of power outages during the storm event. **Greatest negative changes** indicates the extent of the impact in each census tract. The **recovery duration** indicates the length of the impact. Figure 4 shows the distribution of both features of activity density in terms of the income, race, and ethnicity. An activity density of -1 for the greatest negative changes feature indicates that those census tracts recorded very little activity, with the implication being that those census tracts suffered significant power outage impacts. On the other hand, the recovery durations of some census tracts achieved a recovery duration activity density of 20, an assigned value indicating that those census tracts did not return back to the previous normal state before the end of the assessment period in terms of the population activity level.

To examine the difference in impacts on the minority and nonminority populations, this study used the Kruskal-Wallis test, or one-way ANOVA on ranks, due to the non-normality of the residuals. A significant test result indicates that the population median values of the minority and nonminority populations are different. According to the test results of the six pairs of the minority and nonminority (in terms of income, ethnicity and race) in two different assessment features (i.e., the greatest activity density change and recovery duration) shown in Figure 4, the median values of low-income populations in the greatest activity density change and the median values of ethnic minority populations in the recovery duration have considerable differences at the significance level of $p < .1$. On top of that, the median values of low-income populations in greatest activity density change are even significantly different at the significance level of $p < .01$. These results indicate that the low-income populations tended to experience greater impacts, and the ethnic minority groups (i.e., high Hispanic ratio areas) had a longer recovery time from the power outages. Even though there is no statistical difference in the greatest activity density changes between the minority and nonminority groups in race and ethnicity, the median values of the minority groups are both lower than the nonminority groups. In addition, for low-income and racial/ethnic minority groups, the number of significantly impacted census tracts in terms of the greatest changes (i.e., the census tracts with 100% decline in population activity levels) are greater than the number of significantly impacted census tracts for high-income and racial/ethnic nonminority groups. This result implies that more minority census tracts were experiencing significant outages and subsequent impacts than the nonminority census tracts. On the other hand, with respect to recovery duration, the median value for both low-income and racial/ethnic minority categories was six days. In addition, the statistical test results indicate that there is no significant difference in terms of the distribution of income levels and racial minority. Therefore, the next step of the analysis focused on the greatest negative changes in activity density to further reveal the impact of power outages on minority and nonminority groups.

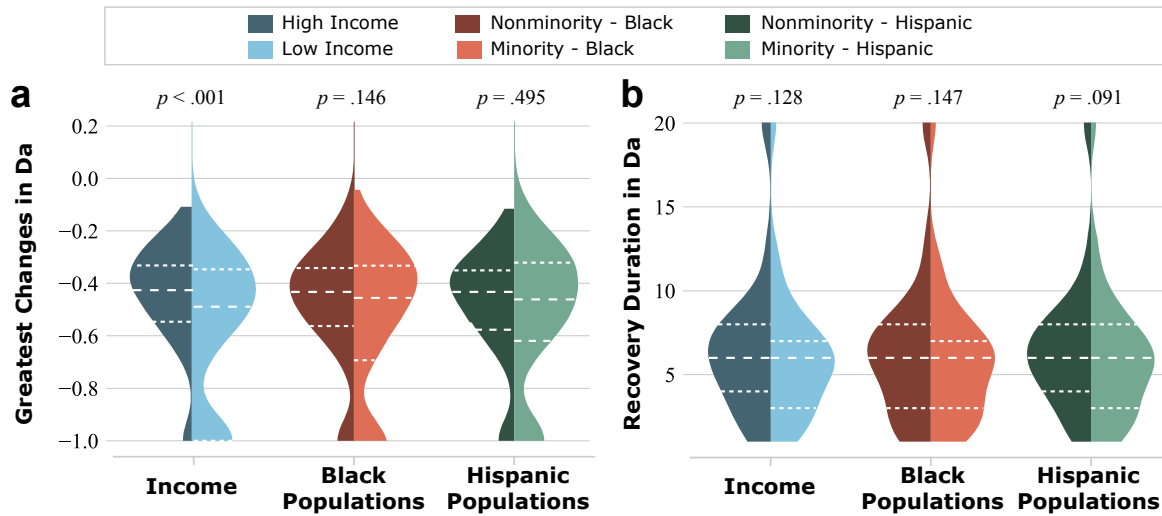

Figure 4 Distribution of the greatest changes (a) and recovery duration (b) in activity density for different income and racial/ethnic groups. Results show that census tracts having more low-income or minority groups were experiencing significant impacts in terms of the extent (more negative changes) and duration (longer recovery duration) of the power outages than census tracts having high-income and nonminority groups.

Of the 786 census tracts in Harris County, and about 13.5% (106) were significantly impacted by power outages, which were identified by very low population activities. Figure 5 depicts the comparison between all census tracts in Harris County and the significantly impacted census tracts from low-income and racial/ethnic minority groups. The median household income of the significantly impacted census tracts is $40,853 USD, which is lower than the $56,429 USD median income of Harris County. The median ratio of persons identifying as Black in the significantly impacted census tracts is 20.43%, which is higher than the 12.01%, the Harris County median. The median ratio of persons identifying as Hispanic in the significantly impacted census tracts is approximately 43.58%, which is higher than the 38.01% the Harris County median. Statistical tests between all census tracts in Harris County and the significantly impacted census tracts from low-income and racial/ethnic minority groups indicate that the median values between the three pairs are all significant at a significance level of $p = .1$. In addition, the p-values of the low-income and racial minority groups are both less than 0.05. Thus, based on the results, the significantly impacted census tracts are likely to include census tracts with a greater proportion of lower-income and racial/ethnic minority residents than in higher-income, nonminority tracts.

These results indicate social inequality in the implementation managed power outage during the Texas winter storm.

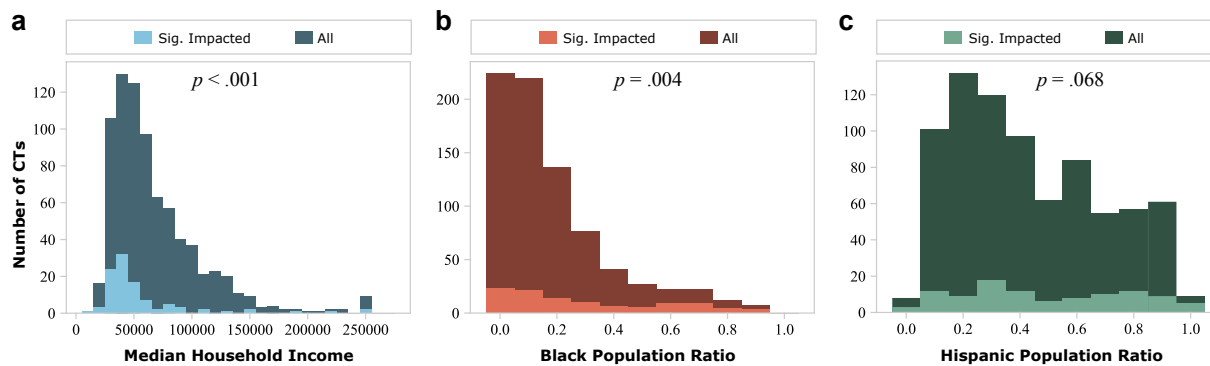

Figure 5 Comparison between all census tracts (CTs) and significantly impacted census tracts due to power outages in Harris County from the aspects of income, race, and ethnicity. The significantly impacted census tracts tend to have lower median income and higher median ratios of Black and Hispanic populations than the medians of Harris County.

Despite the results showing that the higher the percentage of minority population, the more likely a census tract is to be more strongly impacted, exceptions always exist. A census tract with a median household income of $32,285 USD, Hispanic population ratio of 94.15%, and Black population ratio of 4.2% with 30.07% of races other than White and Black population, was not impacted, according to population activity changes. The activity density of this tract was never lower than that of its pre-disaster normal states. In contrast, a census tract with median household income of $226,602 USD, White population of 94.29%, and non-Hispanic population of 95.08%, had about a 40% decrease in activity density, indicating that the census tract experienced impacts due to power outages. Although the fact that the exceptions exist, census tracts having low-income and racial/ethnic minority groups tend to experience greater impacts due to the greater extent and duration of the power outages.

*Impact of Burst Pipes*

To analyze the condition of water services, this study filtered 311 Houston Service Helpline call data on burst pipes based on their service types, such as water leaks, flooding, potable water availability, and poor drainage. As shown in Figure 6, a noticeable spike in 311 water-related case numbers during the impact period of the winter storm can be observed. High volumes of the 311 water-related calls in a census tract may indicate burst-pipe issues due to freezing temperatures. During the impact period of the winter storm, about 67% of the census tracts in Harris County registered water-related 311 cases. Figure 6 also shows the numbers of census tracts receiving the most 311 calls (or case peaks) on each date. Most of the census tracts had case peaks on February 18 and 19, 2021.

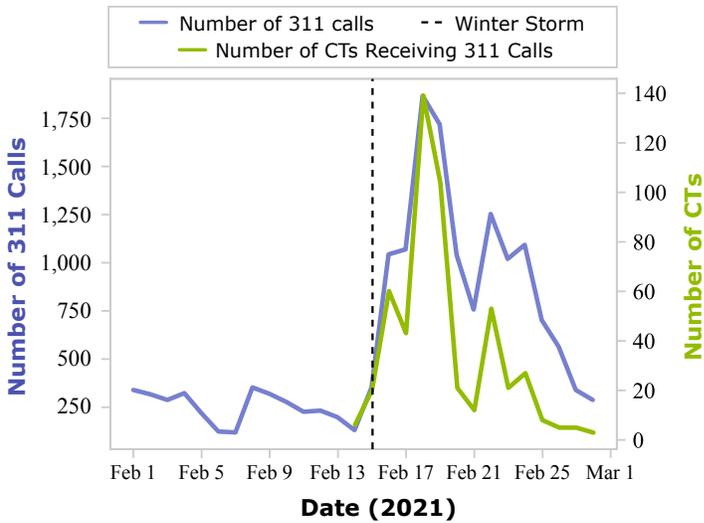

Figure 6 Numbers of the 311 water-related calls (blue) and numbers of census tracts receiving the most water-related calls (orange) during February 2021. Numbers of water-related 311 calls surged after the winter storm reached indicating Harris County underwent burst pipe issues due to freezing temperatures.

Multiple analyses were conducted to understand the impact of burst pipes in minority categories; however, differences in population and area may result in a misunderstanding of results. The greater the population in a census tract, statistically, the more 311 calls from that census tract. Furthermore, a larger census tract area would contain more pipes. To account for the variance among census tracts with respect to population and area, the case peaks of each census tract were normalized by dividing by their population and area. The distribution of the case peaks per area per person for the low-income and racial/ethnic minority groups is shown in Figure 7. The p-values of the statistical test of the differences between the high- and low-income groups and the racial/ethnical minority and nonminority groups are less than 0.05, indicating that the medians of each pair are significantly different. Based on the results, the low-income and racial minority groups had higher normalized case peaks than the high-income and racial nonminority groups. Also, given that a census tract receiving zero 311 calls indicates that a census tract is not impacted by burst pipes, twice as many high-income and racial nonminority census tracts were unimpacted compared with low-income and racial minority groups. This result indicates that low-income and racial minority groups tend to have more 311 calls than the high-income and racial nonminority groups and may suffer more from critical burst pipeline.

A comparison between all census tracts and the impacted census tracts in Harris County was also performed (Figure 8). The comparison for low-income and racial minority groups demonstrates that the differences between all census tracts and impacted census tracts are statistically significant at the significance level of $p = .05$. The median income of Harris County is $56,429 USD, whereas the median income of the impacted census tracts is $50,589 USD. Also, the median percentage of persons identifying as Black within Harris County is 12.01%; in impacted census tracts is, it is 14.13%. Thus, the census tracts impacted by the burst pipes tend to have lower household income and a higher ratio of Black population.

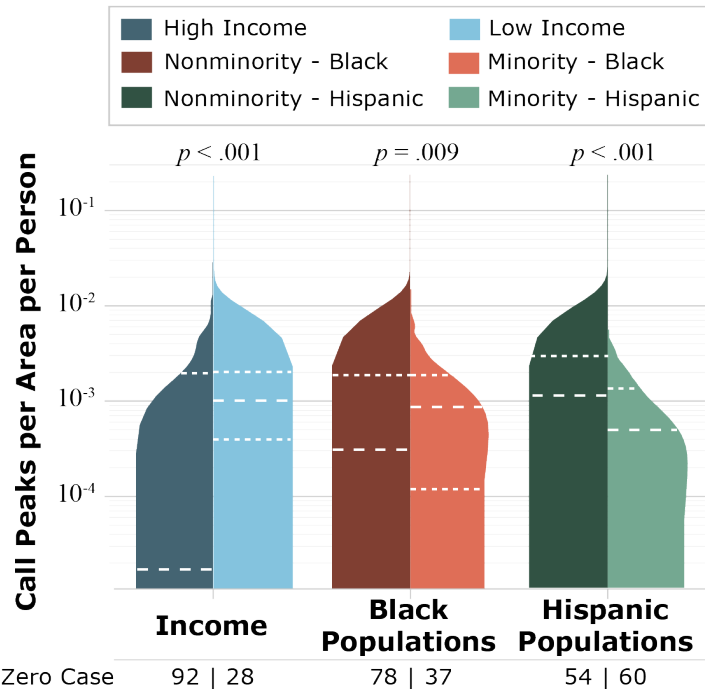

Figure 7 Distribution of the 311 call peaks per area per person for income and race/ethnicity groups and numbers of unimpacted census tracts. Census tracts having more low-income and racial minority groups tend to have higher call peaks per area per person and less unimpacted census tracts.

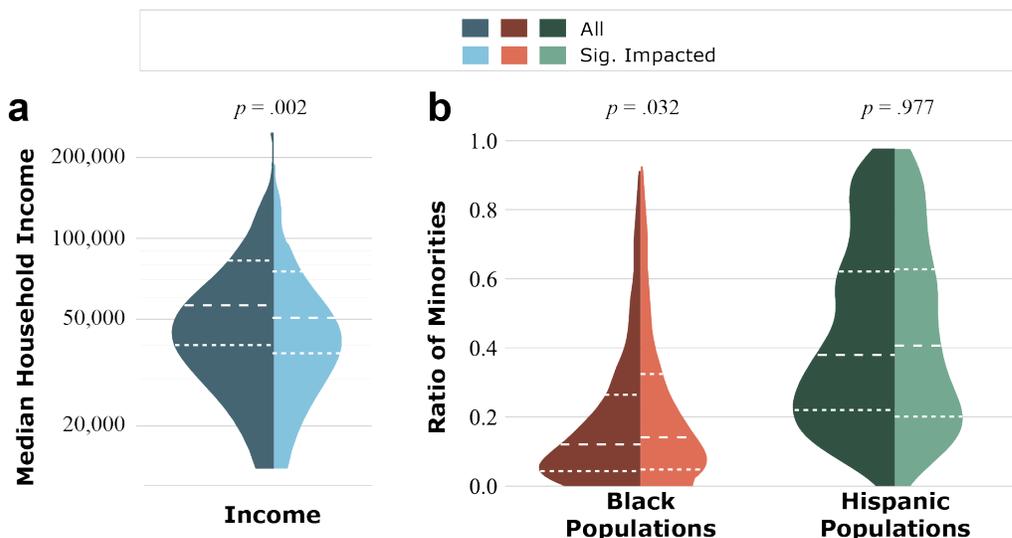

Figure 8 Comparison between all census tracts and the impacted census tracts due to pipe bursts in Harris County from the aspects of income, race, and ethnicity. The impacted census tracts have a lower median income and higher median ratios of Black and Hispanic populations, indicating that the low income and high ratio of Black populations census tracts experienced greater impacts of pipe bursts during the winter storm.

Overall, the 311 data spike indicates that burst pipe issues impacted a large portion of Harris County residents. About 67% of the census tracts requested maintenance on water-related problems. In particular, census tracts with lower-income and higher percentage of Black population are impacted more by burst pipes compared to that with higher-income and lower percentage of Black population. The comparison between all census tracts and the impacted census tracts shows that the impacted census tracts have lower median household income and a higher ratio of Black population. Thus, low income and high percentage of Black population census tracts experienced more burst pipes during the winter storm. This result indicates the disparities in the experience of extreme weather conditions, such as the winter storm, on vulnerable populations (e.g., low-income and racial minority groups).

*Impact of Food Inaccessibility*

In particular, this study obtained data of grocery store and restaurant visits from SafeGraph to reveal fluctuations in food accessibility during the winter storm. Unlike the relationship between activity density and power outages, the food accessibility of a census tract has connections with adjacent census tracts. For example, a resident of one census tract may prefer to buy groceries from a store in a neighboring census tract. Therefore, this study adopted the agglomerative hierarchical clustering algorithm to classify visit trends of census tracts. The advantage of applying this algorithm is that census tracts having similar visit trends can be discovered and grouped to understand the impact of food inaccessibility. Figure 9 shows the classification results of the restaurant and grocery visits in Harris County for every census tract. The census tracts in Harris County are split into four classes, and the trend of each class is represented by the mean of all grouped census tracts. The four classes, from *a* through *d*, represent four patterns of visit trends from the most impacted class to the least impacted class, respectively. An example trend line of class *a* (Figure 9a) is the average of the grouped census tracts (the lighter lines behind trend line of class *a*) and represents the visit fluctuations of class *a*.

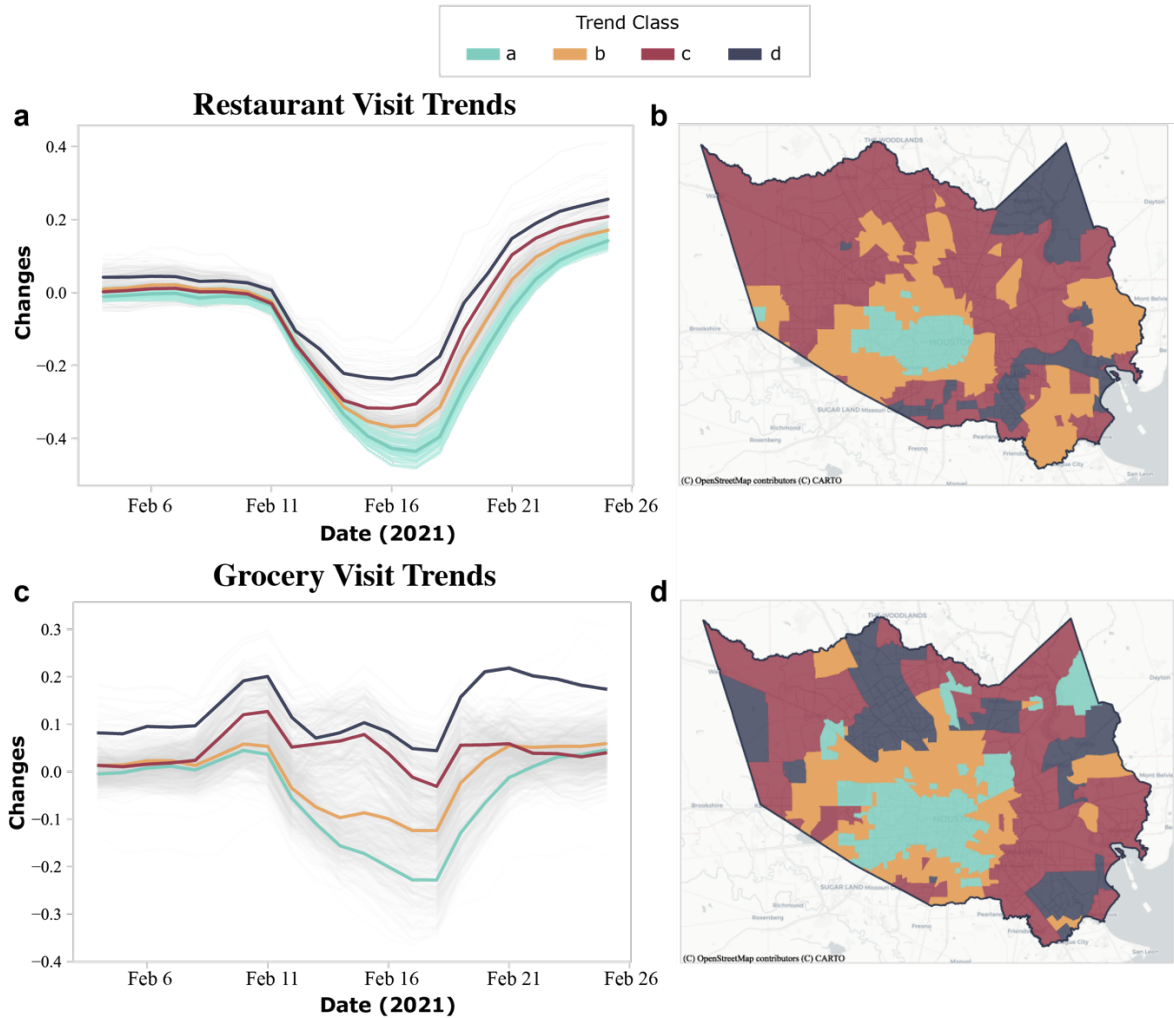

Figure 9 Spatiotemporal distribution of the restaurant (a and b) and grocery store (c and d) visits in Harris County. Four classes, from Class *a* through Class *d*, represent four patterns of visit trends from the most impacted class to the least impacted class, respectively. Trend line is the average of a group of census tracts having similar patterns and represents the visit fluctuations of these census tracts. The most impacted areas (Class *a*) of the restaurant and grocery store visits are close to Houston's downtown areas.

The study classified different visit trends for the restaurants and grocery stores in Harris County and presented the spatial distribution of each visit trend in Figure 9. For both types of POIs, class *a* represents the most impacted census tracts in terms of visitor numbers. In contrast, class *d* shows the least impacted census tracts. For the restaurant visit trends, the shapes of the four classes are almost identical except for the magnitude of the impact level (Figure 9a). Visit trends started to decrease on February 11, 2021 and showed the most significant dips February 16 through February 17, 2021. On the other hand, the shapes of the grocery store visit trends (Figure 9c) are not perfectly identical but demonstrate similar patterns during the winter storm. The grocery store visits had one major and one minor increase before the storm (February 15, 2021), which shows the preparedness for the winter storm. The largest dips in visits for the grocery stores were on February 18, 2021, and they returned to previous normal states afterward. Overall, the locations of the most impacted areas of both grocery store and restaurant points of interest are close to Houston's downtown areas, which is to be expected since the downtown areas are where people work rather than where they live. Due to the weather and traffic, a population would tend to avoid traveling to work and thus lead to fewer visits to business areas compared to the other areas. In addition, most of the secondary impacted areas (class *b*) are census tracts that surround census tracts identified as class *a* (Figure 9b and Figure 9d). By evaluating the spatial distribution of the visit trends, the most impacted regions in terms of food accessibility can be identified in areas surrounding the inner part of Harris County.

In the next step, the relationships between food accessibility and the three sub-population categories were evaluated. Subpopulation categories may experience food inaccessibility. For example, when public transportation services were suspended, a low-income household who cannot afford a self-owned vehicle might be unable to travel to grocery stores due to a lack of viable transportation. Figure 10 shows the relationship between the four food accessibility classes and the three minority categories. The most impacted census tracts (census tracts identified as class *a*) regarding restaurant and grocery store visits tend to have larger nonminority populations than the other three classes (classes *b*, *c*, and *d*). This result may be due to the fact that these census tracts are primarily located in the downtown area of Houston, which attracts a more affluent population. Due to a reduced downtown workforce during the storm, visits to restaurants and grocery stores experienced decreased. On the other hand, the median income for the secondary impacted census tracts (census tracts identified as class *b*) is relatively lower than the other three classes (classes *a*, *c*, and *d*), and the median ratio of Black and Hispanic populations for the secondary impacted census tracts (census tracts identified as class *b*) is somewhat higher than the other three classes (classes *a*, *c*, and *d*). This result indicates that those census tracts classified as the secondary impacted classes have a lower income and a high ratio of Black and Hispanic populations. Therefore, the census tracts with low-income and racial/ethnic minority groups were likely to have food accessibility issues.

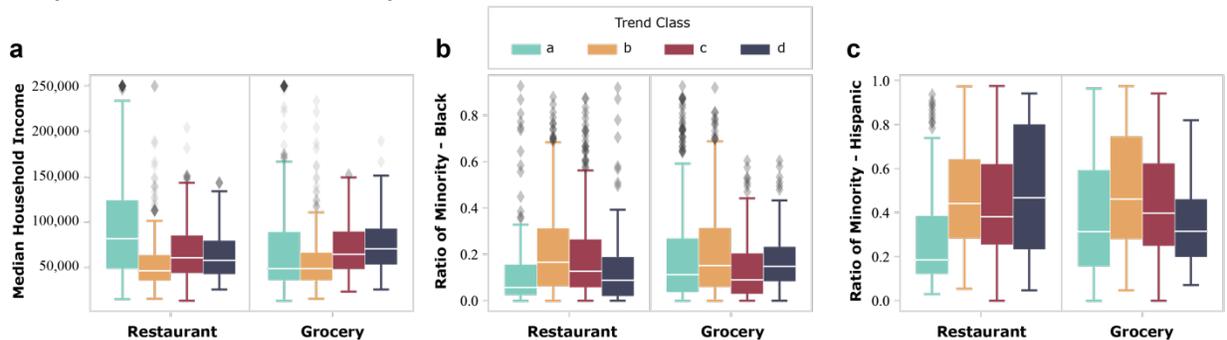

Figure 10 Relationships between food accessibility classes and income and racial/ethnic minority groups. Results show that the secondary impacted census tracts generally have lower median income and higher median ratios of Black and Hispanic populations.

## Discussions and Closing Remarks

In this study, community-scale big data from various sources were used to quantify and assess the spatial patterns of the responses and impacts of the winter storm. The quantified impacts of managed power outages, burst pipes, and food accessibility were then applied to examine the extent of disparities in impacts experienced by low-income and racial/ethnic minority subpopulations. This study used the impact of Texas Winter Storm Uri in February 2021, an unusual extreme weather event, on Harris County (Texas) as a case study. Critical infrastructures, such as power-generating equipment, highways, and the water distribution system, were not well-prepared to withstand a winter storm of Uri's magnitude. During this critical situation, power-generation plant operators were forced to interrupt services to avoid severely damaging the entire power grid. For example, because power demand increased due to the load imposed by home heating, and power supply decreased due to unwinterized power equipment, ERCOT needed to shut down partial power and implement rotating power outages to protect the power grid. It is vital to understand the extent to which these managed power outages were done in an equitable way for different subpopulations. To assess such potential disparate impacts, authors made a strong effort to obtain granular and high-resolution power outage data by contacting respectable organizations. Yet much data was not publicly available at the time the authors analyzed the issues. In this case, lack of publicly available data provides an opportunity to demonstrate the value of using this community-scale big data for rapid impact assessments. Given that these data become available in the future, results could be further verified and discussed.

Results in all three aspects—impacts of power outages, pipe bursts, and food accessibility—indicate the disparities in the experience of extreme weather conditions on vulnerable populations. For power outages and food accessibility, results show that census tracts having larger percentages low-income and racial/ethnic minority groups tend to have been more strongly disrupted by the winter storm. Also, the impact assessment results of burst pipes demonstrate that low-income and racial minority groups are more heavily disrupted than high-income and racial nonminority groups. Impacts of power outages in this winter storm indicate how infrastructure operators respond to an unusual event. Even though ruggedization of power equipment cannot be accomplished in a short time, the infrastructure operator can make plans for extreme weather scenarios within the considerations of social vulnerability prior to a disaster and inform residents about predetermined hours or days that each area could expect outages when a disaster reaches. In this way, people can be better prepared. Impacts of burst pipes indicate limited awareness and preparedness of infrastructure owners and residents. Given that temperatures below freezing were predicted, actions can be taken to protect pipes. Impacts of food inaccessibility indicate the influences of service disruptions of critical infrastructures, such as public transportation, road networks, grocery stores, and restaurants.

The disparate impacts revealed in this study can raise awareness in infrastructure managers to take equitable resilience seriously. Service providers should prepare mitigation plans for emergency response and account for these revealed disparities. Rapid assessment results in this study demonstrate that community-scale big data can serve as a tool for local agencies to understand the impact of past severe weather events so that a more equitable allocation of resources can be planned before the arrival of next disaster.

**Data availability**

The data that support the findings of this study are available from Mapbox, SafeGraph, and 311, but restrictions apply to the availability of these data, which were used under license for the current study. The data can be accessed upon request submitted on Mapbox and SafeGraph. Other data we use in this study are all publicly available.

# Code availability

The code that supports the findings of this study is available from the corresponding author upon request.

# References

1. Coronese, M., F. Lamperti, K. Keller, F. Chiaromonte, and A. Roventini. Evidence for Sharp Increase in the Economic Damages of Extreme Natural Disasters. *Proceedings of the National Academy of Sciences*, Vol. 116, No. 43, 2019, pp. 21450–21455. https://doi.org/10.1073/pnas.1907826116.
2. Kryvasheyeu, Y., H. Chen, N. Obradovich, E. Moro, P. Van Hentenryck, J. Fowler, and M. Cebrian. Rapid Assessment of Disaster Damage Using Social Media Activity. *Science Advances*, Vol. 2, No. 3, 2016, p. e1500779. https://doi.org/10.1126/sciadv.1500779.
3. Stocker, T., Ed. *Climate Change 2013: The Physical Science Basis: Working Group I Contribution to the Fifth Assessment Report of the Intergovernmental Panel on Climate Change*. Cambridge University Press, New York, 2014.
4. Fan, C., and A. Mostafavi. A Graph-Based Method for Social Sensing of Infrastructure Disruptions in Disasters. *Computer-Aided Civil and Infrastructure Engineering*, Vol. 34, No. 12, 2019, pp. 1055–1070. https://doi.org/10.1111/mice.12457.
5. Kim, J., and M. Hastak. Online Human Behaviors on Social Media During Disaster Responses. *The Journal of the NPS Center for Homeland Defense and Security*, 2017.
6. Lu, L., X. Wang, Y. Ouyang, J. Roningen, N. Myers, and G. Calfas. Vulnerability of Interdependent Urban Infrastructure Networks: Equilibrium after Failure Propagation and Cascading Impacts: Vulnerability of Interdependent Urban Infrastructure Networks. *Computer-Aided Civil and Infrastructure Engineering*, Vol. 33, No. 4, 2018, pp. 300–315. https://doi.org/10.1111/mice.12347.
7. Panteli, M., and P. Mancarella. Influence of Extreme Weather and Climate Change on the Resilience of Power Systems: Impacts and Possible Mitigation Strategies. *Electric Power Systems Research*, Vol. 127, 2015, pp. 259–270. https://doi.org/10.1016/j.epsr.2015.06.012.
8. Fothergill, A., E. G. M. Maestas, and J. D. Darlington. Race, Ethnicity and Disasters in the United States: A Review of the Literature. *Disasters*, Vol. 23, No. 2, 1999, pp. 156–173. https://doi.org/10.1111/1467-7717.00111.
9. Donner, W., and H. Rodríguez. Population Composition, Migration and Inequality: The Influence of Demographic Changes on Disaster Risk and Vulnerability. *Social Forces*, Vol. 87, No. 2, 2008, pp. 1089–1114. https://doi.org/10.1353/sof.0.0141.
10. Hong, B., B. J. Bonczak, A. Gupta, and C. E. Kontokosta. Measuring Inequality in Community Resilience to Natural Disasters Using Large-Scale Mobility Data. *Nature Communications*, Vol. 12, No. 1, 2021, p. 1870. https://doi.org/10.1038/s41467-021-22160-w.
11. Morss, R. E., O. V. Wilhelmi, G. A. Meehl, and L. Dilling. Improving Societal Outcomes of Extreme Weather in a Changing Climate: An Integrated Perspective. *Annual Review of Environment and Resources*, Vol. 36, No. 1, 2011, pp. 1–25. https://doi.org/10.1146/annurev-environ-060809-100145.

12. Bolin, B., and L. C. Kurtz. Race, Class, Ethnicity, and Disaster Vulnerability. In *Handbook of Disaster Research* (H. Rodríguez, W. Donner, and J. E. Trainor, eds.), Springer International Publishing, Cham, pp. 181–203.
13. Elliott, J. R., and J. Pais. Race, Class, and Hurricane Katrina: Social Differences in Human Responses to Disaster. *Social Science Research*, Vol. 35, No. 2, 2006, pp. 295–321. https://doi.org/10.1016/j.ssresearch.2006.02.003.
14. Reid, M. Disasters and Social Inequalities. *Sociology Compass*, Vol. 7, No. 11, 2013, pp. 984–997. https://doi.org/10.1111/soc4.12080.
15. Peacock, W. G., S. V. Zandt, Y. Zhang, and W. E. Highfield. Inequities in Long-Term Housing Recovery After Disasters. *Journal of the American Planning Association*, Vol. 80, No. 4, 2014, pp. 356–371. https://doi.org/10.1080/01944363.2014.980440.
16. Mostafavi, A. A System-of-Systems Framework for Exploratory Analysis of Climate Change Impacts on Civil Infrastructure Resilience. *Sustainable and Resilient Infrastructure*, Vol. 3, No. 4, 2018, pp. 175–192. https://doi.org/10.1080/23789689.2017.1416845.
17. Coleman, N., A. Esmalian, and A. Mostafavi. Equitable Resilience in Infrastructure Systems: Empirical Assessment of Disparities in Hardship Experiences of Vulnerable Populations during Service Disruptions. *Natural Hazards Review*, Vol. 21, No. 4, 2020, p. 04020034. https://doi.org/10.1061/(ASCE)NH.1527-6996.0000401.
18. Podesta, C., N. Coleman, A. Esmalian, F. Yuan, and A. Mostafavi. Quantifying Community Resilience Based on Fluctuations in Visits to Points-of-Interest Derived from Digital Trace Data. *Journal of The Royal Society Interface*, Vol. 18, No. 177, 2021, p. rsif.2021.0158, 20210158. https://doi.org/10.1098/rsif.2021.0158.
19. Zhang, C., W. Yao, Y. Yang, R. Huang, and A. Mostafavi. Semiautomated Social Media Analytics for Sensing Societal Impacts Due to Community Disruptions during Disasters. *Computer-Aided Civil and Infrastructure Engineering*, Vol. 35, No. 12, 2020, pp. 1331–1348. https://doi.org/10.1111/mice.12576.
20. Li, Q., Z. Tang, N. Coleman, and A. Mostafavi. Detecting Early-Warning Signals in Time Series of Visits to Points of Interest to Examine Population Response to COVID-19 Pandemic. *IEEE Access*, Vol. 9, 2021, pp. 27189–27200. https://doi.org/10.1109/ACCESS.2021.3058568.
21. Fan, C., C. Zhang, A. Yahja, and A. Mostafavi. Disaster City Digital Twin: A Vision for Integrating Artificial and Human Intelligence for Disaster Management. *International Journal of Information Management*, Vol. 56, 2021, p. 102049. https://doi.org/10.1016/j.ijinfomgt.2019.102049.
22. Resch, B., F. Usländer, and C. Havas. Combining Machine-Learning Topic Models and Spatiotemporal Analysis of Social Media Data for Disaster Footprint and Damage Assessment. *Cartography and Geographic Information Science*, Vol. 45, No. 4, 2018, pp. 362–376. https://doi.org/10.1080/15230406.2017.1356242.
23. Ragini, J. R., P. M. R. Anand, and V. Bhaskar. Big Data Analytics for Disaster Response and Recovery through Sentiment Analysis. *International Journal of Information Management*, Vol. 42, 2018, pp. 13–24. https://doi.org/10.1016/j.ijinfomgt.2018.05.004.
24. Fan, C., F. Wu, and A. Mostafavi. A Hybrid Machine Learning Pipeline for Automated Mapping of Events and Locations From Social Media in Disasters. *IEEE Access*, Vol. 8, 2020, pp. 10478–10490. https://doi.org/10.1109/ACCESS.2020.2965550.

25. Xiao, Y., Q. Huang, and K. Wu. Understanding Social Media Data for Disaster Management. *Natural Hazards*, Vol. 79, No. 3, 2015, pp. 1663–1679. https://doi.org/10.1007/s11069-015-1918-0.
26. Fan, C., M. Esparza, J. Dargin, F. Wu, B. Oztekin, and A. Mostafavi. Spatial Biases in Crowdsourced Data: Social Media Content Attention Concentrates on Populous Areas in Disasters. *Computers, Environment and Urban Systems*, Vol. 83, 2020, p. 101514. https://doi.org/10.1016/j.compenvurbsys.2020.101514.
27. Morstatter, F., J. Pfeffer, H. Liu, and K. Carley. Is the Sample Good Enough? Comparing Data from Twitter's Streaming API with Twitter's Firehose. *Proceedings of the International AAAI Conference on Web and Social Media*, Vol. 7, No. 1, 2013.
28. Li, Q., L. Bessell, X. Xiao, C. Fan, X. Gao, and A. Mostafavi. Disparate Patterns of Movements and Visits to Points of Interest Located in Urban Hotspots across US Metropolitan Cities during COVID-19. *Royal Society Open Science*, Vol. 8, No. 1, 2021, p. 201209. https://doi.org/10.1098/rsos.201209.
29. Juhasz, L., and H. Hochmair. Studying Spatial and Temporal Visitation Patterns of Points of Interest Using SafeGraph Data in Florida. *GIS Center*, 2020. https://doi.org/10.1553/giscience2020_01_s119.
30. Yuan, F., Y. Yang, Q. Li, and A. Mostafavi. Unraveling the Temporal Importance of Community-Scale Human Activity Features for Rapid Assessment of Flood Impacts. *arXiv:2106.08370 [physics]*, 2021.
31. Gao, X., C. Fan, Y. Yang, S. Lee, Q. Li, M. Maron, and A. Mostafavi. Early Indicators of Human Activity During COVID-19 Period Using Digital Trace Data of Population Activities. *Frontiers in Built Environment*, Vol. 6, 2021. https://doi.org/10.3389/fbuil.2020.607961.
32. Almasy, S., H. Silverman, and J. Sutton. More than 150 Million Americans under Winter Weather Alerts as Record Cold Temps Make Life Miserable. *CNN*, February 16, 2021.
33. Freedman, A., J. Muyskens, and J. Samenow. Central States' Arctic Plunge: The Historic Cold Snap and Snow by the Numbers. *Washington Post*, February 24, 2021.
34. Mulcahy, S. At Least 111 People Died in Texas during Winter Storm, Most from Hypothermia. *The Texas Tribune*, March 25, 2021.
35. Texas Health and Human Services News Updates: Winter Storm-Related Deaths. https://web.archive.org/web/20210526151612/https://dshs.texas.gov/news/updates.shtm#wn. Accessed June 11, 2021.
36. Cappucci, M. 133-Car Pileup on Fort Worth Highway during Freezing Rain Leaves at Least 6 Dead. *Washington Post*, February 11, 2021.
37. Ferman, M. Winter Storm Could Cost Texas More Money than Any Disaster in State History. *The Texas Tribune*, February 25, 2021.
38. Watson, K. P., R. Cross, M. P. Jones, G. Buttorff, P. Pinto, S. L. Sipole, and A. Vallejo. *The Effects of the Winter Storm of 2021 in Harris County*. Hobby School of Public Affairs, University of Houston.
39. Hastie, trevor, R. Tibshirani, and J. Friedman. *The Elements of Statistical Learning: Data Mining, Inference, and Prediction*. Springer.
40. McCullough, E. D., Kate McGee and Jolie. Texas Leaders Failed to Heed Warnings That Left the State's Power Grid Vulnerable to Winter Extremes, Experts Say. *The Texas Tribune*. https://www.texastribune.org/2021/02/17/texas-power-grid-failures/. Accessed July 28, 2021.